# UNIFIED ANALYTICAL SOLUTION FOR RADIAL FLOW TO A WELL IN A CONFINED AQUIFER

**Phoolendra Kumar Mishra and Velimir V. Vesselinov**

Computational Earth Science Group

Earth and Environmental Sciences Division

Los Alamos National Laboratory

MS T003, Los Alamos, NM 87544 USA

## ABSTRACT

Drawdowns generated by extracting water from a large diameter (e.g. water supply) well are affected by wellbore storage. We present an analytical solution in Laplace transformed space for drawdown in a uniform anisotropic aquifer caused by withdrawing water at a constant rate from a partially penetrating well with storage. The solution is back transformed into the time domain numerically. When the pumping well is fully penetrating our solution reduces to that of *Papadopulos and Cooper* [1967]; *Hantush* [1964] when the pumping well has no wellbore storage; *Theis* [1935] when both conditions are fulfilled and *Yang et.al. [2006]* when the pumping well is partially penetrating, has finite radius but lacks storage. We use our solution to explore graphically the effects of partial penetration, wellbore storage and anisotropy on time evolutions of drawdown in the pumping well and in observation wells.

## INTRODUCTION

When water is pumped from a large diameter (e.g. water supply) well drawdown in the surrounding aquifer is affected by temporal decline in wellbore storage. An analytical solution accounting for this effect under radial flow toward a fully penetrating well of finite diameter with storage was developed by *Papadopulos and Cooper* [1967]. A corresponding solution without wellbore storage was presented earlier by *van Everdingen and Hurst* [1949] and later, in elliptical coordinates, by *Kucuk and Brigham* [1979]. *Mathias and Butler* [2007] extended the solution of *Kucuk and Brigham* [1979] by adding wellbore storage and horizontal anisotropy. Their solution utilized Mathieu functions in Laplace transformed space and numerical inversion of the result into the time domain. *Yang et.al.* [2006] extended the solution of *van Everdingen and Hurst* [1949] by allowing the pumping well to be partially penetrating. *Dougherty and Babu* [1984] developed an analytical solution for a pumping well with storage in a confined double porosity reservoir. Their solution can be reduced to that for a single porosity confined aquifer but ignores anisotropy. None of the available analytical solutions account simultaneously for aquifer anisotropy, partial penetration and storage capacity of the pumping well under confined aquifer conditions.

*Moench* [1997, 1998] developed an analytical solution for flow to a pumping well with storage in a uniform anisotropic unconfined (water table) aquifer. We present a new solution for radial flow to a partially penetrating well of finite diameter with storage in an anisotropic confined aquifer. Whereas *Moench* [1997, 1998] used Fourier cosine series in Laplace transformed space we employ Laplace transformation with respect to time followed by finite cosine transformation with respect to vertical coordinates. Our solution reduces to that of *Papadopulos and Cooper* [1967] when the pumping well is fully penetrating, *Hantush* [1964] in

43  the absence of wellbore storage, *Theis* [1935] when both conditions are fulfilled, and *Yang et.al.*
44  [2006] when the pumping well is partially penetrating, has finite radius but lacks storage. We
45  use our solution to explore graphically the effects of partial penetration, wellbore storage and
46  anisotropy on time evolutions of drawdown in the pumping well and in observation wells.

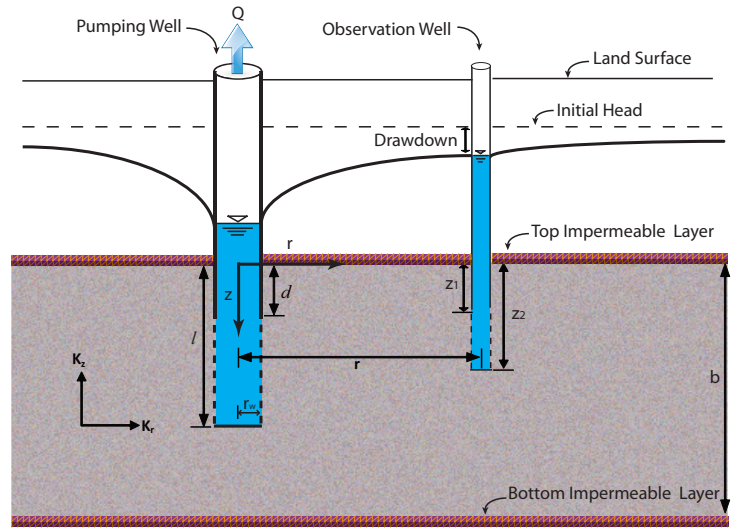

47

48  **Figure 1:** Schematic representation of system geometry

49  **THEORY**

50  **Problem Definition**

51  Consider a well of finite radius $r_w$ that is in hydraulic contact with a surrounding
52  confined aquifer at depths $d$ through $l$ below the top (Figure 1). The aquifer is horizontal and
53  of infinite lateral extent with uniform thickness $b$, uniform hydraulic properties and anisotropy
54  ratio $K_D = K_z / K_r$ between vertical and horizontal hydraulic conductivities, $K_z$ and $K_r$,
55  respectively. Initially, drawdown $s(r,z,t)$ throughout the aquifer is zero where $r$ is radial
56  distance from the axis of the well, $z$ is depth below the top of the aquifer and $t$ is time. Starting

57   at time $t = 0$ water is withdrawn from the pumping well at a constant volumetric rate $Q$.

58   Consider the bottom of the well to be impermeable and ignore flow beneath it. Then drawdown

59   distribution in space-time is controlled by

60   $$K_r\left(\frac{\partial^2 s}{\partial r^2} + \frac{1}{r}\frac{\partial s}{\partial r}\right) + K_z\frac{\partial^2 s}{\partial z^2} = S_s\frac{\partial s}{\partial t} \qquad 0 \leq z < b \qquad (1)$$

61   subject to

62   $$s(r,z,0) = s(\infty,z,t) = 0 \qquad\qquad r \geq r_w \qquad (2)$$

63   $$\frac{\partial s}{\partial z} = 0 \quad \text{at } z = 0 \ \& \ z = b \qquad\qquad r > r_w$$

64   (3)

65   $$2\pi(l-d)K_r r_w \left(\frac{\partial s}{\partial r}\right)_{r=r_w} - C_w\left(\frac{\partial s}{\partial t}\right)_{r=r_w} = -Q \qquad d < z < l \qquad (4)$$

66   $$r\left(\frac{\partial s}{\partial r}\right) = 0 \qquad\qquad 0 < z < d \ \& \ l < z < b \qquad (5)$$

67   where $C_w$ is wellbore storage coefficient (volume of water released from well storage per unit

68   drawdown in it).

69   **Solution in Laplace Space**

70       We show in Appendix A that the Laplace transform of the solution, indicated by an

71   overbar, is given by

$$\bar{s}(r_D, z_D, p_D) = \frac{Qt}{4\pi K_r b} \left\{ \frac{2}{p_D} \frac{K_0(\phi_0)}{r_{wD}\phi_0 K_1(r_{wD}\phi_0) + \frac{C_{wD}}{2(l_D - d_D)} r_{wD}^2 \phi_0^2 K_0(r_{wD}\phi_0)} \right.$$

(6)

$$\left. + \frac{4}{p_D \pi (l_D - d_D)} \sum_{n=1}^{\infty} \frac{K_0(\phi_n) [\sin(n\pi l_D) - \sin(n\pi d_D)] \cos(n\pi z_D)/n}{r_{wD}\phi_n K_1(r_{wD}\phi_n) + \frac{C_{wD}}{2(l_D - d_D)} r_{wD}^2 \phi_0^2 K_0(r_{wD}\phi_n)} \right\}$$

where $r_D = r/b$, $z_D = z/b$, $p_D = pt$, $r_{wD} = r_w/r$, $C_{wD} = C_w/(\pi S_s r_w^2)$, $d_D = d/b$, $l_D = l/b$, $\phi_n = \sqrt{p_D/t_s + \beta^2 n^2 \pi^2}$, $t_s = \alpha_s t/r^2$, $\alpha_s = K_r/S_s$ and $\beta = r_D K_D^{1/2}$, $K_0$ and $K_1$ being modified Bessel functions of second kind and order zero and one, respectively. A corresponding solution in the time domain $s(r_D, z_D, p_D)$, is obtained through numerical inversion of the Laplace transform by means of an algorithm due to *Crump* [1976] as modified by *de Hoog et. al.* [1982]. Whereas standard inversion with respect to $p$ is done over a time interval $[0,t]$, we do the inversion with respect to $p_D$ over a unit dimensionless time (corresponding to $p_D^{-1}$) interval $[0,1]$, regardless of what $t_s$ is.

**Vertically Averaged Drawdown**

Drawdown in a piezometer or observation well that penetrates the aquifer between dimensionless depths $z_{D1} = z_1/b$ and $z_{D2} = z_2/b$ at a dimensionless radial distance $r_D$ from the pumping well (Figure 1) is obtained by averaging the point drawdown over this interval according to

$$\bar{s}_{z_{D2} - z_{D1}}(r_D, p_D) = \frac{1}{z_{D2} - z_{D1}} \int_{z_{D1}}^{z_{D2}} s(r_D, z_D, p_D) dz_D.$$

(7)

88  Substituting (6) into (7) and evaluating the integral gives

$$\overline{s}_{z_{D2}-z_{D1}}(r_D, p_D) = \frac{Qt}{4\pi K_r b} \left\{ \frac{2}{p_D} \frac{K_0(\phi_0)}{r_{wD}\phi_0 K_1(r_{wD}\phi_0) + \frac{C_{wD}}{2(l_D - d_D)} r_{wD}^2 \phi_0^2 K_0(r_{wD}\phi_0)} \right.$$

$$\left. + \frac{4}{p_D \pi^2 (l_D - d_D)(z_{D2} - z_{D1})} \sum_{n=1}^{\infty} \frac{K_0(\phi_n)\left[\sin(n\pi l_D) - \sin(n\pi d_D)\right]\left[\sin(n\pi z_{D2}) - \sin(n\pi z_{D1})\right]/n^2}{r_{wD}\phi_n K_1(r_{wD}\phi_n) + \frac{C_{wD}}{2(l_D - d_D)} r_{wD}^2 \phi_0^2 K_0(r_{wD}\phi_n)} \right\} \quad (8)$$

### Reduction to Solution of *Papadopoulos and Cooper* [1967]

91  When the pumping well is fully penetrating $l_D = 1$, $d_D = 0$ and (6) reduces to the
92  corresponding Laplace domain solution of *Papadopoulos and Cooper* [1967],

$$\overline{s}(r_D, p_D) = \frac{Qt}{4\pi K_r b} \left\{ \frac{2}{p_D} \frac{K_0\left(\sqrt{\frac{p_D}{t_s}}\right)}{r_{wD}\sqrt{\frac{p_D}{t_s}} K_1\left(r_{wD}\sqrt{\frac{p_D}{t_s}}\right) + \frac{C_{wD}}{2} r_{wD}^2 \frac{p_D}{t_s} K_0\left(r_{wD}\sqrt{\frac{p_D}{t_s}}\right)} \right\} \quad (9)$$

### Reduction to Solution's of *Yang et.al.* [2006], *Hantush* [1964] and *Theis* [1935]

95  When the pumping well has finite diameter $(r_w \neq 0)$ but negligible or no wellbore
96  storage $(C_{wD} \to 0)$, (6) reduces to the solution of *Yang et. al.* [2006] in Laplace space,

$$\overline{s}(r_D, z_D, p_D) = \frac{Qt}{4\pi K_r b} \left\{ \frac{2}{p_D} \frac{K_0(\phi_0)}{r_{wD}\phi_0 K_1(r_{wD}\phi_0)} \right.$$

$$\left. + \frac{4}{p_D \pi (l_D - d_D)} \sum_{n=1}^{\infty} \frac{K_0(\phi_n)\left[\sin(n\pi l_D) - \sin(n\pi d_D)\right]\cos(n\pi z_D)/n}{r_{wD}\phi_n K_1(r_{wD}\phi_n)} \right\} \quad (10)$$

98  When the pumping well has small diameter $(r_w \to 0)$, (6) reduces to *Hantush*'s [1964]
99  solution in Laplace space due to the fact that $xK_1(x) \to 1$ and $x^2 K_0(x) \to 0$ as $x \to 0$,

100 $$\bar{s}(r_D, z_D, p_D) = \frac{Qt}{4\pi K_r b} \left\{ \frac{2}{p_D} K_0\left(\sqrt{\frac{p_D}{t_s}}\right) + \frac{4}{p_D \pi} \sum_{n=1}^{\infty} \frac{K_0(\phi)\left[\sin(n\pi l_D) - \sin(n\pi d_D)\right]\cos(n\pi z_D)}{n(l_D - d_D)} \right\}$$

101 . (11)

102 It is well established and easily verified that the latter in turn reduces to the *Theis* [1935]

103 solution in Laplace space when the pumping well becomes fully penetrating $(d_D = 0, l_D = 1)$,

104 $$\bar{s}(r_D, p_D) = \frac{Qt}{4\pi K_r b} \left\{ \frac{2}{p_D} K_0\left(\sqrt{\frac{p_D}{t_s}}\right) \right\}$$ (12)

105 ## 3. RESULTS AND DISCUSSION

106 To investigate the effect of partial penetration, wellbore storage and anisotropy on

107 drawdown we consider a pumping well of dimensionless radius $r_w/b = 0.02$.

108 **Drawdown in pumping well**

109 We start by considering drawdown in a pumping well penetrating the upper half

110 ($d_D = 0.0, l_D = 0.5$) of an isotropic aquifer with $K_D = 1.0$. Figure 2 compares the variation of

111 dimensionless drawdown $s_D(r_D, z_D, t_s) = (4\pi K_r b/Q)s(r_D, z_D, t_s)$ in the pumping well with

112 dimensionless time $t_s$ using different analytical solutions when $C_{wD} = 1.0 \times 10^2$. At early time

113 water is derived entirely from wellbore storage, rendering dimensionless drawdown linearly

114 proportional to dimensionless time (forming a line with unit slope on log-log scale); our

115 solution and that of *Papadopulos and Cooper [1967]* reflect this clearly. Solutions that do not

116 account for wellbore storage predict a much earlier rise in drawdown. Whereas the *Papadopulos*

117 *and Cooper [1967]* solution approaches that of *Theis* [1935] at later dimensionless time, ours

118 approaches that of *Hantush* [1964] as the effects of finite radius and wellbore storage dissipate.

119 The solution of *Yang et al.* [2006], which considers only the first effect, exhibits an earlier rise in dimensionless drawdown than do any of the other solutions, eventually coinciding with that of *Hantush* [1964]. Dimensionless drawdown in the pumping well at late dimensionless time exceeds that predicted by solutions which ignore partial penetration.

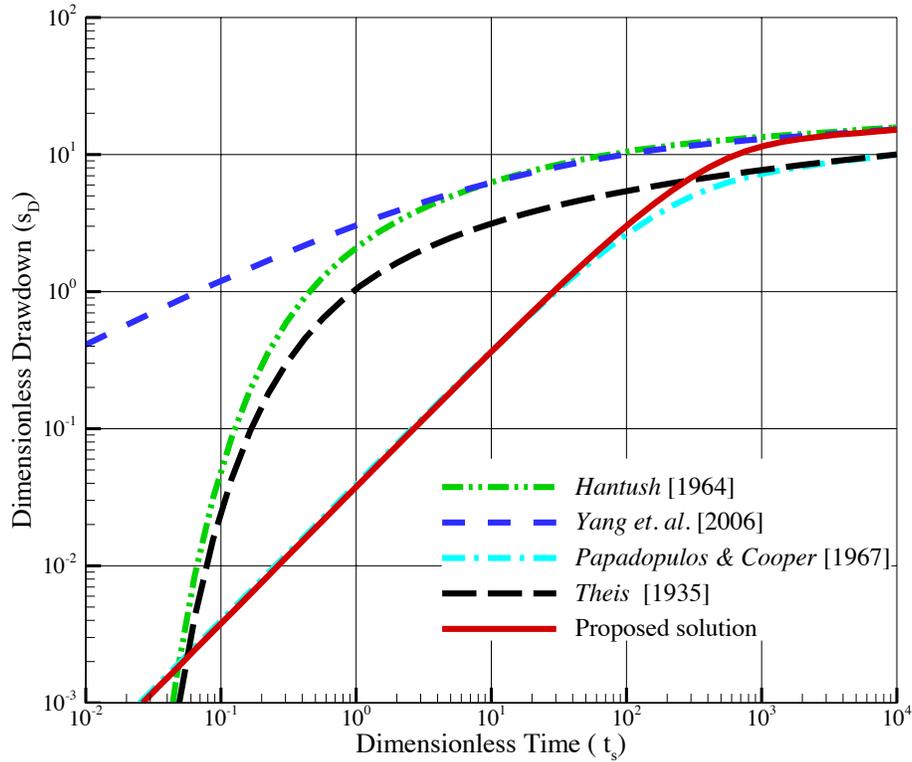

**Figure 2:** Dimensionless drawdown in pumping well versus dimensionless time, computed by various analytical solutions when $C_{wD} = 1.0 \times 10^2$, $d_D = 0.0$, $l_D = 0.5$ and $K_D = 1.0$.

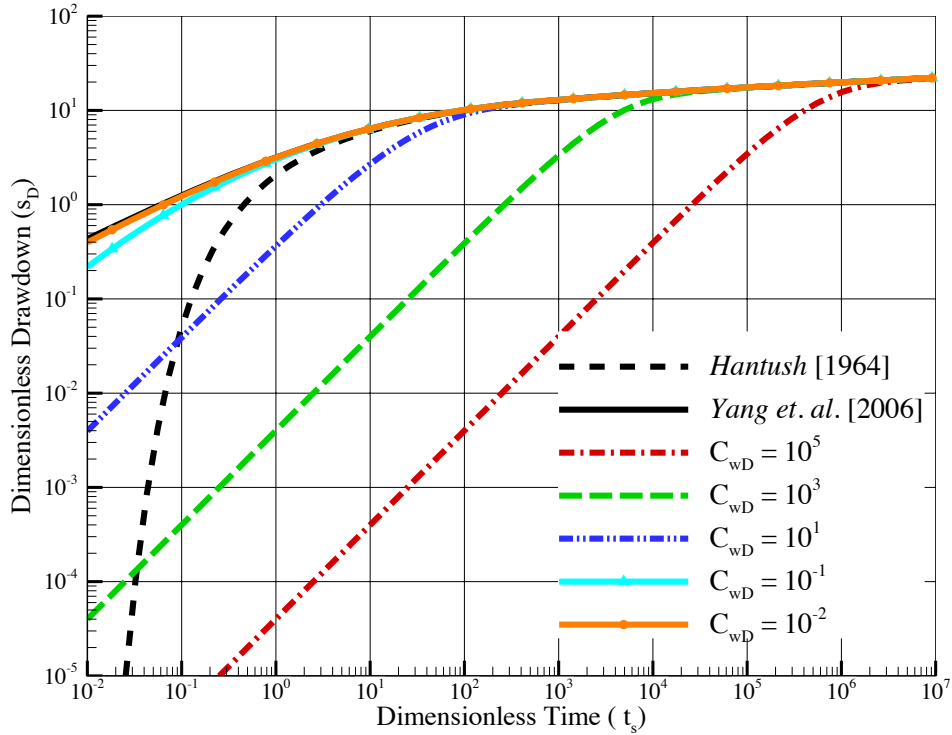

126

127 **Figure 3:** Dimensionless drawdown in pumping well versus dimensionless time for various

128 values of dimensionless wellbore storage $C_{wD}$ when $d_D = 0.0$, $l_D = 0.5$ and $K_D = 1.0$.

129 Figure 3 shows how dimensionless drawdown in the pumping well varies with

130 dimensionless time $t_s$ for different values of the dimensionless wellbore storage coefficient,

131 $C_{wD}$. As with the solution of *Papadopulos and Cooper* [1967], the larger is $C_{wD}$ the longer does

132 wellbore storage impact drawdown in the pumping well. As $C_{wD}$ diminishes our solution

133 approaches that of *Yang et.al* [2006].

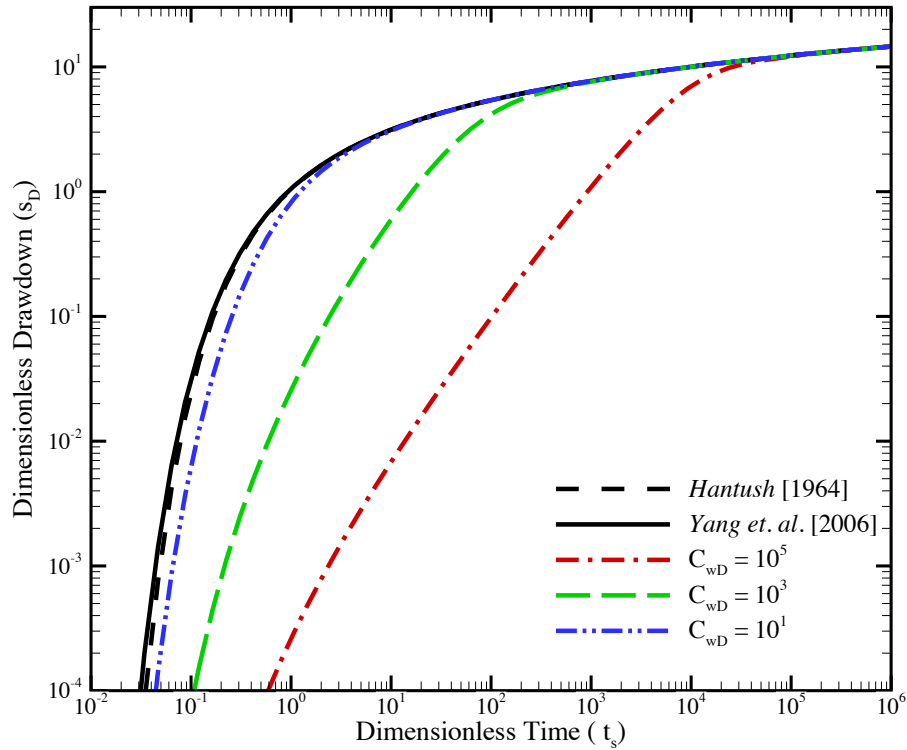

**Figure 4:** Dimensionless drawdown at $z_D = 0.5$ and $r_D = 0.2$ versus dimensionless time for various values of dimensionless wellbore storage $C_{wD}$ when $l_D = 0.5$ and $K_D = 1.0$.

**Drawdown in piezometer**

Figure 4 shows dimensionless time-drawdown variations at dimensionless radial distance $r_D = 0.2$ from the axis of the pumping well and dimensionless elevation $z_D = 0.5$ (midway between the horizontal no-flow boundaries) for different values of $C_{wD}$ under the above conditions. When $C_{wD}$ is large, the early dimensionless time-drawdown curve on log-log scale is nearly linear with a unit slope, reflecting a strong effect of storage in the pumping well on early drawdown in a nearby piezometer. As $C_{wD}$ diminishes this effect becomes less discernible, the curve becoming nonlinear and steeper. The curve tends asymptotically toward the solution of *Yang et al.* [2006], which in turn is very close to that of *Hantush* [1964] due to the small dimensionless radius we have assigned to the pumping well in our example.

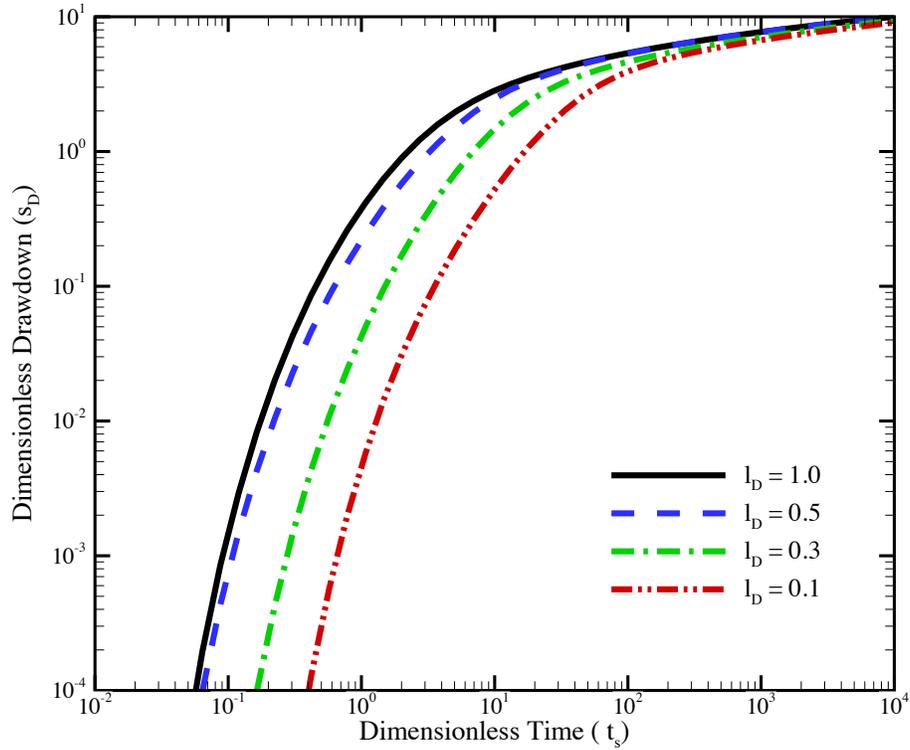

**Figure 5:** Dimensionless drawdown at $z_D = 0.5$ and $r_D = 0.2$ versus dimensionless time for various screen lengths $l_D$ when $C_{wD} = 1.0 \times 10^2$, $d_D = 0$ and $K_D = 1.0$.

Figure 5, corresponding to the case where $C_{wD} = 1.0 \times 10^2$, shows that dimensionless drawdown at $r_D = 0.2$ and $z_D = 0.5$ increases when the pumping well is extended to the aquifer bottom ($d_D = 0.0, l_D = 1.0$ below the observation point) but decreases when this well becomes shallower; a similar trend is reflected in the solution of *Hantush* [1964]. Reducing the ratio $K_D$ between vertical and horizontal hydraulic conductivity in the case of a well that is shallower than the observation point ($d_D = 0.0, l_D = 0.25$) likewise causes dimensionless drawdown at this point to diminish (Figure 6).

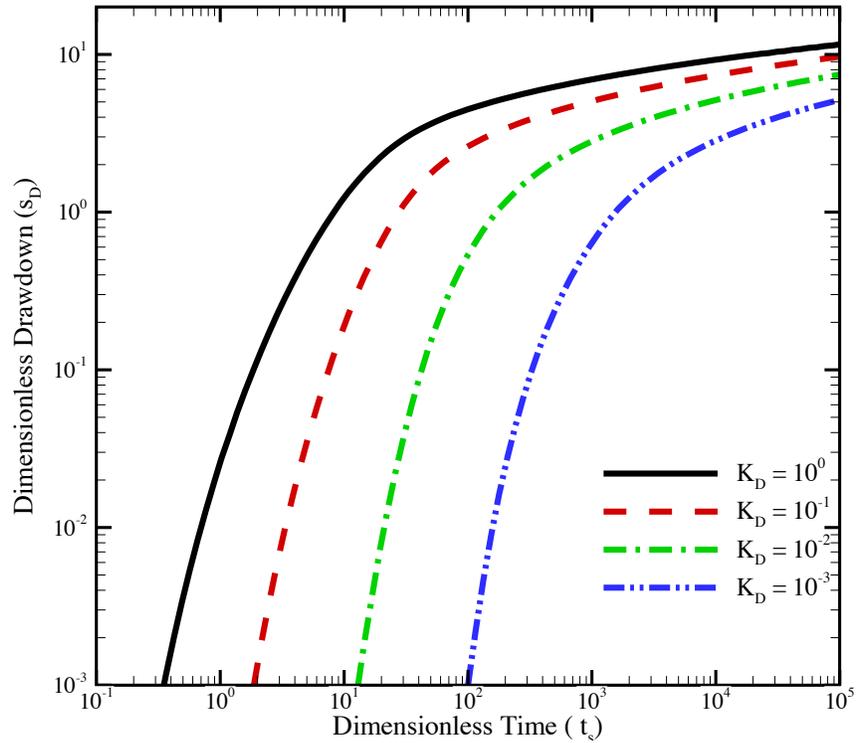

**Figure 6:** Dimensionless drawdown at $z_D = 0.5$ and $r_D = 0.2$ versus dimensionless time for various anisotropy ratios $K_D = K_z / K_r$ when $C_{wD} = 1.0 \times 10^2$, $d_D = 0.0$ and $l_D = 0.25$.

Figure 7 illustrates the impact of dimensionless radial distance from the pumping well on dimensionless time-drawdown at $z_D = 0.5$ when $d_D = 0.0$, $l_D = 0.25$, $K_D = 1$ and $C_{wD} = 1.0 \times 10^2$. As this distance increases the effects of both wellbore storage and partial penetration diminish, the dimensionless time-drawdown response in the aquifer approaching that predicted by *Theis* [1935].

## 4. SUMMARY AND CONCLUSION

A new analytical solution has been developed for a partially penetrating well of finite diameter with storage pumping at a constant rate from an anisotropic confined aquifer. Our solution unifies the solutions of *Papadopulos and Cooper* [1967], *Hantush* [1964], *Theis*

[1935] and *Yang et.al.* [2006] by accounting simultaneously for aquifer anisotropy, partial penetration and wellbore storage capacity of the pumping well under confined conditions. We used our solution to explore all three effects. Reducing the anisotropy ratio $K_D = K_z/K_r$ causes drawdown in the aquifer to decrease. Whereas the effect of partial penetration decreases with increasing distance from the pumping well, that of wellbore storage diminishes with distance and time.

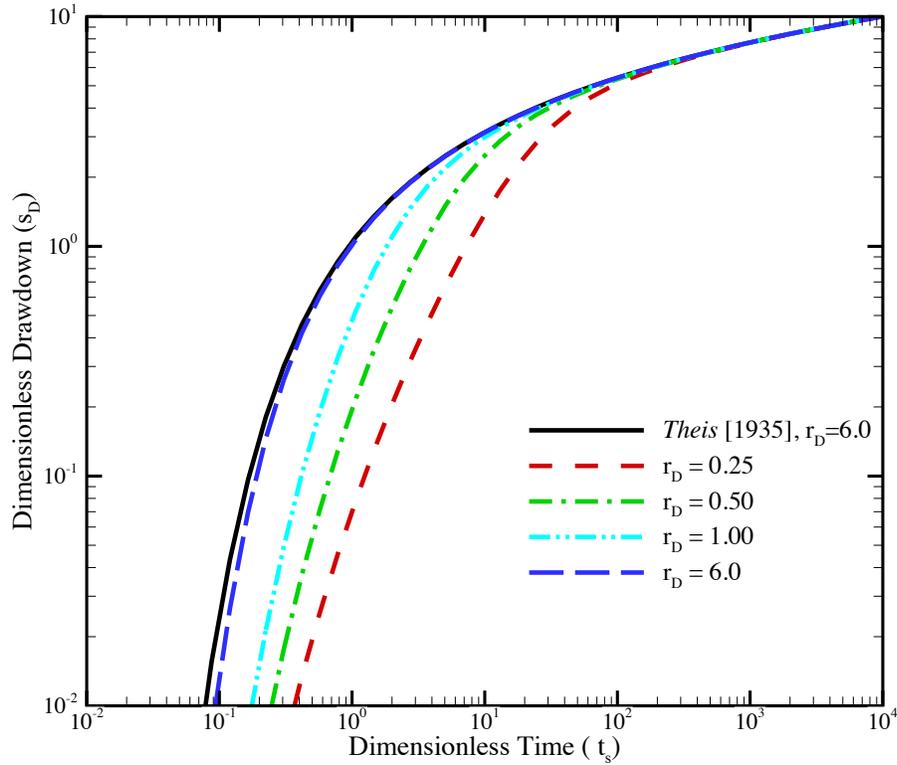

**Figure 7:** Dimensionless drawdown versus dimensionless time at $z_D = 0.5$ and various values of $r_D = r/b$ when $C_{wD} = 1.0 \times 10^3$, $d_D = 0.0$, $l_D = 0.25$ and $K_D = 1.0$.

**Appendix A: Laplace transformed drawdown**

Introducing a new variable $r' = r(K_z/K_r)^{1/2} = rK_D^{1/2}$ and taking Laplace transform of (1) – (5) gives

181 $$\frac{\partial^2 \bar{s}}{\partial r'^2} + \frac{1}{r'}\frac{\partial \bar{s}}{\partial r'} + \frac{\partial^2 \bar{s}}{\partial z^2} = \frac{S_s}{K_z} p\bar{s} \qquad 0 \le z < b \qquad (A1)$$

182

183 subject to

184 $$\bar{s}(\infty, z, p) = 0 \qquad (A2)$$

185

186 $$\frac{\partial \bar{s}}{\partial z} = 0 \quad \text{at} \quad z = 0 \ \& \ z = b \qquad r > r_w \qquad (A3)$$

187 $$2\pi(l-d)K_r r'_w \left(\frac{\partial \bar{s}}{\partial r'}\right)_{r'=r'_w} - C_w p \bar{s}_{r'=r'_w} = -\frac{Q}{p} \qquad d < z < l \qquad (A4)$$

188 $$r'_w \left(\frac{\partial \bar{s}}{\partial r'}\right)_{r'=r'_w} = 0 \qquad 0 < z < d \ \& \ l < z < b \qquad (A5)$$

189

190 Defining the finite cosine transform of $\bar{s}(r', z, p)$ as (*Churchill*, 1958, p.354-355)

191 $$f_c\{\bar{s}(r',z,p)\} = \bar{s}_c(r',n,p) = \int_0^b \bar{s}(r',z,p)\cos(n\pi z/b)\,dz \qquad n = 0,1,2,... \qquad (A6)$$

192 with inverse

193 $$\bar{s}(r',z,p) = \frac{1}{b}\bar{s}_c(r',0,p) + \frac{2}{b}\sum_{n=1}^{\infty}\bar{s}_c(r',n,p)\cos(n\pi z/b) \qquad (A7)$$

194 implies that, by virtue of (A3),

195 $$f_c\left\{\frac{\partial^2 \bar{s}}{\partial z^2}\right\} = -\left(\frac{n\pi}{b}\right)^2 \bar{s}_c(r',n,p) + (-1)^n \left.\frac{\partial \bar{s}(r',z,p)}{\partial z}\right|_{z=b} - \left.\frac{\partial \bar{s}(r',z,p)}{\partial z}\right|_{z=0} = -\left(\frac{n\pi}{b}\right)^2 \bar{s}_c(r',n,p)$$

196 (A7)

197 Hence finite cosine transformation of (A1) – (A5) leads to

198 $$\frac{\partial^2 \bar{s}_c}{\partial r'^2} + \frac{1}{r'}\frac{\partial \bar{s}_c}{\partial r'} - \left[\frac{p}{K_z/S_s} + \left(\frac{n\pi}{b}\right)^2\right]\bar{s}_c = 0 \qquad (A8)$$

199 $$\bar{s}_c(\infty, n, p) = 0 \qquad (A9)$$

200 $$2\pi(l-d)K_r r'_w \left(\frac{\partial \bar{s}_c}{\partial r'}\right)_{r'=r'_w} - pC_w(\bar{s}_c)_{r'=r'_w} = -\frac{Q}{p}\int_d^l \cos(n\pi z/b)dz$$
$$= -\frac{Q}{p}(b/n\pi)\left[\sin(n\pi l/b) - \sin(n\pi d/b)\right]$$

201 (A10)

202 The general solution of (A8) is

203 $$\bar{s}_c(r', n, p) = AK_0(Nr') + BI_0(Nr') \qquad (A11)$$

204 where $N^2 = \dfrac{p}{K_z/S_s} + (n\pi/b)^2$, $I_0$ and $K_0$ being modified Bessel functions of first and second

205 kind, respectively, and of zero order. By virtue of (A9) $B = 0$. Substituting this and (A11) into

206 (A10), noting that $\partial K_0(Nr')/\partial r' = -NK_1(Nr')$, solving for $A$ and substituting back into (A11) yields

207 $$\bar{s}_c(r', n, p) = \frac{Q}{p}\frac{(b/n\pi)\left[\sin(n\pi l/b) - \sin(n\pi d/b)\right]}{2\pi(l-d)K_r Nr'_w K_1(Nr'_w) + pC_w K_0(Nr'_w)}K_0(Nr'). \qquad (A12)$$

208 Noting that $\lim\limits_{n\to 0}\left[l\dfrac{\sin(n\pi l/b)}{n\pi l/b} - d\dfrac{\sin(n\pi d/b)}{n\pi d/b}\right] = l - d$ one gets

209 $$\bar{s}_c(r', 0, p) = \frac{Q}{p}\frac{K_0\left(r'\sqrt{\dfrac{p}{K_z/S_s}}\right)}{2\pi K_r r'_w \sqrt{\dfrac{p}{K_z/S_s}}K_1\left(r'_w\sqrt{\dfrac{p}{K_z/S_s}}\right) + \dfrac{pC_w}{(l-d)}K_0\left(r'_w\sqrt{\dfrac{p}{K_z/S_s}}\right)}. \qquad (A13)$$

210 This allows obtaining the inverse Fourier cosine transform of (A12),

$$\bar{s}(r',z,p) = \frac{1}{b}\frac{Q}{K_r p} \frac{K_0\left(r'\sqrt{\dfrac{p}{K_z/S_s}}\right)}{2\pi r'_w \sqrt{\dfrac{p}{K_z/S_s}} K_1\left(r'_w \sqrt{\dfrac{p}{K_z/S_s}}\right) + \dfrac{pC_w}{K_r(l-d)} K_0\left(r'_w \sqrt{\dfrac{p}{K_z/S_s}}\right)}$$
$$+ \frac{2}{b}\frac{Q}{K_r p} \sum_{n=1}^{\infty} \frac{(b/n\pi)[\sin(n\pi l/b) - \sin(n\pi d/b)]}{2\pi(l-d)Nr'_w K_1(Nr'_w) + \dfrac{pC_w}{K_r} K_0(Nr'_w)} \cos(n\pi z/b) K_0(Nr') \quad (A14)$$

Recalling that $r' = rK_D^{1/2}$ and rewriting (A14) in dimensionless form yields (6).

## REFERENCES

Churchil, R.V. (1958), *Operational Mathematics*, 2nd edition, McGraw Hill, New York.

Crump, K. S. (1976), Numerical inversion of Laplace transforms using a Fourier series approximation, *J. Assoc. Comput. Mach.*, vol.*23*, issue 1, 89–96.

de Hoog, F. R., J. H. Knight, and A. N. Stokes (1982), An improved method for numerical inversion of Laplace transforms, *SIAM J. Sci. Stat. Comput.*, vol. *3 Issue 3*, 357–366, doi:10.1137/0903022.

Dougherty, D. E., and D. K. Babu (1984), Flow to a partially penetrating well in a double-porosity reservoir, Water Resour. Res., 20, 1116- 1122.

Hantush, M. S. (1964), Hydraulics of wells, *Adv. Hydrosci.*, *1*, 281–442.

Kucuk, F., and W. E. Brigham (1979), Transient flow in elliptical systems, *Soc. Pet, Eng. J.*, 19, 401-410, doi:10.2118/7488-PA.

Mathias, S. A., and A. P. Butler (2007), Flow to a finite diameter well in a horizontally anisotropic aquifer with wellbore storage, *Water Resour. Res.*, *43*, W07501, doi: 10.1029/2006WR005839.

Moench, A. F. (1997), Flow to a well of finite diameter in a homogenous anisotropic water table aquifer, *Water Resour. Res.*, 33, No.6, 1397- 1407.

Moench, A. F. (1998), Correction to "Flow to a well of finite diameter in a homogenous anisotropic water table aquifer", *Water Resour. Res.*, 34, No.9, 2431-2432.

Papadopulos, I. S., and H. H. Cooper Jr. (1967), Drawdown in a well of large diameter, *Water Resour. Res.*, *3*(1), 241–244.

Theis, C. V. (1935), The relationship between the lowering of the piezometric surface and rate and duration of discharge of a well using groundwater storage, *Eos Trans. AGU*, *16*, 519–524.

Van Everdingen, A.F. and Hurst, W. (1949), The application of the Laplace transformation to flow problems in reservoirs, *Trans AIME*, 186, 305-324.

Yang, S.-Y., H.-D. Yeh, and P.-Y. Chiu (2006), A closed form solution for constant flux pumping in a well under partial penetration condition, *Water Resour. Res.*, 42, W05502, doi:10.1029/2004WR003889.